\documentclass{iau}

\usepackage{amsmath}
\usepackage{graphicx}
\usepackage{multirow}

\begin{document}

\lefttitle{Andrea Ercolino}
\righttitle{Progenitor models of supernovae interacting with their binary companions}

\jnlPage{1}{7}
\jnlDoiYr{2026}
\doival{10.1017/xxxxx}

\aopheadtitle{Proceedings IAU Symposium}
\editors{The editors}

\title{Progenitor models of supernovae interacting with their binary companions}

\author{Andrea Ercolino}
\affiliation{Argelander Institut für Astronomie, Auf dem Hügel 71, 53121 Bonn, Germany  \email{aercolino@astro.uni-bonn.de}}

\begin{abstract}
The appearance of nearly all core-collapse (CC) supernovae (SNe) is largely affected by the interaction between their progenitors and a close binary companion. Using a comprehensive, state-of-the-art grid of single- and binary stellar evolution models, the relative frequencies of SN types are predicted, as well as the distribution functions of their main properties. SNe that interact with nearby circumstellar material are included, namely those that explode during ongoing binary mass transfer, and are found to account for $\sim5\%$ of all CCSNe, consistently with observations. The interaction between the newly born compact object and the companion is also investigated, which may produce observable signatures similar to those seen in SN2022jli. These SNe could account for between $3\%$ and $27\%$ of all H-poor CCSNe. These results will help develop strategies for identifying such supernovae in past and future searches, which will help to constrain uncertain physics in single and binary evolution models.
\end{abstract}


\maketitle

\section{Introduction}
Stars initially more massive than $\sim9\,\mathrm{M}_\odot$ typically explode as core-collapse (CC) supernovae (SNe).  These explosions affect the evolution of their host galaxies through both dynamical and chemical feedback. Understanding these explosive transients requires having a comprehensive picture of the evolutionary pathways and physical properties of their progenitor stars, as they influence the resulting observable signatures of the SNe and their rates.

Recent observations show that most massive stars are born with a close binary companion \citep{Sana_massive_stars_binaries, Sana25_binaries_bloem}, with whom they are expected to transfer mass during their lives. This substantially alters the pre-explosion evolution of SN progenitors relative to single stars: some stars lose or gain large amounts of mass as donors or accretors, while others merge, producing even more unusual structures.

The classification scheme of CCSNe \citep[see, e.g.,][]{GalYam_Handbook} provides clues to the properties of their progenitors. The presence (lack) of strong hydrogen lines in the spectra defines Type~IIP/L (Type~Ibc) SNe, which are thought to arise from H-rich (H-poor) stars which underwent limited (significant) mass loss. SNe with weak or short-lived hydrogen lines are classified as Type~IIb, indicating progenitors that lost most, but not all, of their H-rich envelopes. As such, the relative populations of different CCSN types are a key probe of mass loss in massive stars. \cite{Smith2011_obsSN} showed that the high fraction of stripped-envelope SNe and the brightest observed Type~II SNe could not be simultaneously explained by single-star evolution, implying that binary evolution likely plays a major role in producing stripped-envelope SNe.


The past decade has seen an explosion of wide-field transient surveys that have increased the quantity and detail of observed transients, offering the best data yet on the relative populations of different SN types \citep[e.g.,][]{Graur2017, Pessi2025, Srivastav2026}. In parallel, advances in detailed binary evolution calculation have enabled the systematic production of large-scale model grids \citep[e.g.,][]{Marchant2017,Wang2020, Gilkis2025, Jin2026}. This paper summarizes the main results from \cite{Ercolino2026a,Ercolino2026b}, which use state-of-the-art binary evolution models to predict the relative fractions and population-wide features of CCSNe and compare them against recent results from transient surveys.

\section{Methods}
A grid-based binary population synthesis code, \texttt{SN-ORACLE} \citep[][]{Ercolino2026a, Ercolino2026b}, is used to postprocess the large-scale grid of state-of-the-art detailed binary
evolution models from \cite{Jin2026}, together with single-star models computed by \cite{Jin2024} and \cite{Ercolino2026a}. The resulting predictions include the relative rates of different CCSN types and population-wide distributions of progenitor and explosion properties. 

The underlying binary grid spans a wide range of birth parameters. In particular, it covers initial masses in the range $5\leq M_\mathrm{i}/M_\odot\leq 100$, making it effectively complete for CCSN progenitors, and it also covers binaries with initial orbital periods of $1.1\leq P_\mathrm{i}/\,\mathrm{d}\leq 5000$, from systems that merge at birth to those so wide as to never interact. Each binary and single-star evolution model is assigned a birth probability, by adopting a birth-binary fraction of $75\%$ \citep{Sana_massive_stars_binaries}, a \cite{Salpeter1955} initial mass-function, as well as flat initial mass-ratio and log-flat initial orbital period distributions.  

The adoption of detailed binary evolution models favors physical detail over extensive exploration of model uncertainties, but several model assumptions can still be varied, yielding different population models for different choices of assumptions. In particular, different assumptions are made on i) the distribution of natal kicks imparted to newborn neutron stars, which determine whether binaries remain bound after the initially more massive star explodes, thus shaping the later evolution of the initially less massive companion, ii) the onset of unstable mass transfer, using the detailed mass transfer history to determine whether the system merges, and iii) whether a star explodes as a SN or directly collapses into a black hole, based on its structure at CC. For exploding models, the structure at CC, the composition of the ejecta, and recent mass-loss history are used to assign the SN type (see \citealt{Ercolino2026a} for details).

\section{The role of binary interaction on supernova progenitors} 

\begin{figure}
    \centering
    \includegraphics[width=\linewidth]{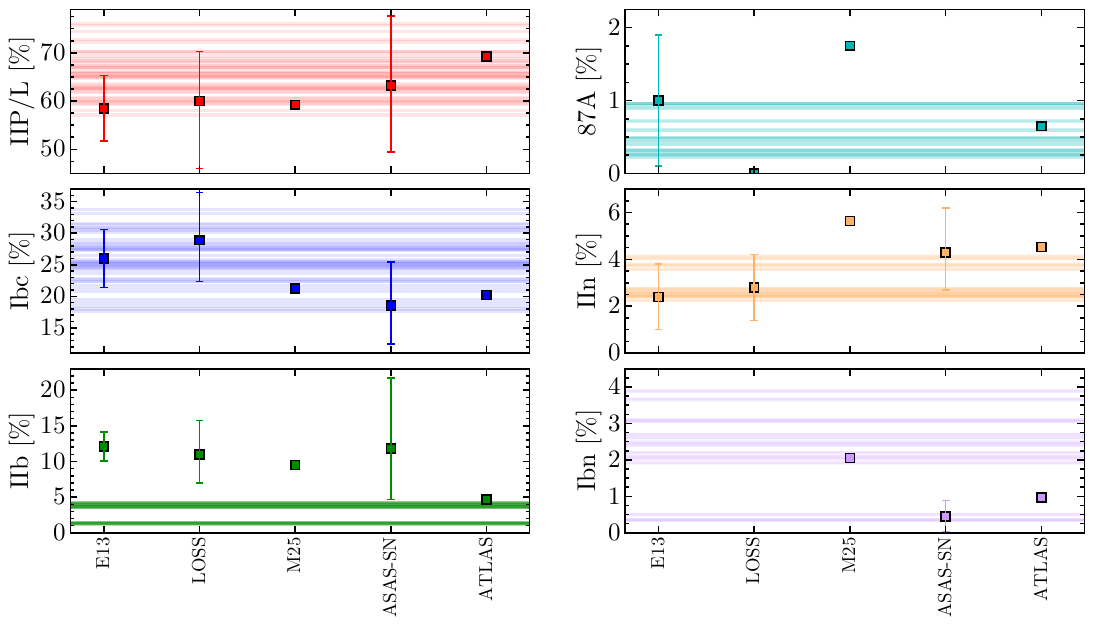}
    \caption{Fraction of CCSNe of different types (different panels) predicted across different population models (horizontal bands) compared to the values inferred from volume-limited samples (scatter) from \citet[E13]{Eldridge2013}, \cite{Graur2017} from the LOSS survey's high-mass galaxy subsample (LOSS), \citet[M25]{Ma2025}, \cite{Pessi2025} for the ASAS-SN survey (ASAS-SN), and \cite{Srivastav2026} for the ATLAS100 survey (ATLAS). Figure adapted from \cite{Ercolino2026a}.}
    \label{fig:SN_pop}
\end{figure}

Across all population models, Type~IIP/L SNe make up the majority of CCSNe, around $57-76\%$, broadly consistent with observationally inferred rates (Fig.\,\ref{fig:SN_pop}). The majority of their progenitors are affected by binary interaction ($56-68\%$), more specifically as accretors or merger products. This means that less than half of the progenitors of Type~IIP/L SNe come from effectively single-stars, which are the models typically considered in many calculations. Consequently, the properties of the resulting SNe may be more diverse than those currently modeled when using single-star progenitor models.

\begin{figure}
    \centering
    \includegraphics[width=0.7\linewidth]{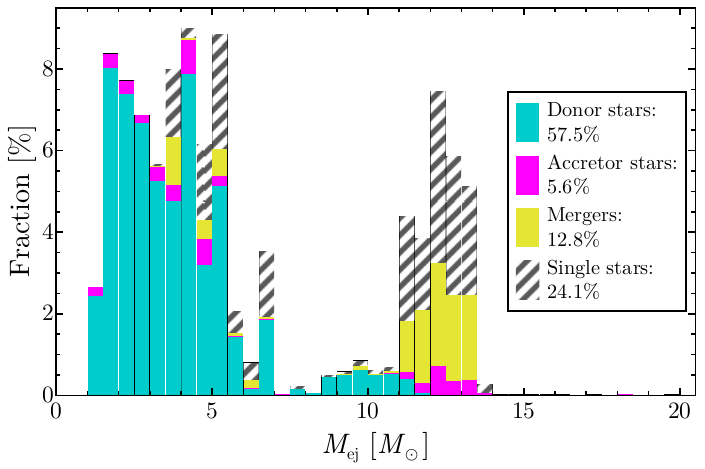}
    \caption{Ejecta mass distribution of Type~Ibc SNe in one population model.  Different colors denote different progenitors, from donor stars in binaries (cyan), accretors in binaries (magenta), mergers (yellow) and effectively single-stars (hatched). The legend reports the relative contribution of each progenitor to Type~Ibc SNe. Figure adapted from \cite{Ercolino2026a}.}
    \label{fig:Mej_Ibc}
\end{figure}

Type~Ibc SNe are predicted to make up around $18-34\%$ of all CCSNe, consistent with observations (Fig.\,\ref{fig:SN_pop}). Most of their progenitors ($70$–$97\%$) have undergone binary evolution, mainly as mass donors. Many population models predict a bimodal ejecta-mass distribution (Fig.\,\ref{fig:Mej_Ibc}), with a low-mass peak dominated by donor stars that lost their envelopes through binary mass transfer and a high-mass peak dominated by accretors, mergers, and effectively single stars that lost their H-rich envelopes through winds (see \citealt{Ercolino2026a} for details). The high-mass peak depends sensitively on the criteria used to determine whether the progenitor explodes after CC, and in some cases it vanishes entirely. Observations suggest the existence of Type~Ibc SNe with both low-mass \citep{Das2024} and high-mass ejecta \citep{Taddia2019, K19}. However, a volume-limited sample is not yet available for consistent population-wide comparisons.

Type~IIb SNe are predicted to make up around $1-4\%$ of all CCSNe. They seem to be underproduced, compared to the observationally inferred rate of about $5-22\%$ (Fig.\,\ref{fig:SN_pop}). This could be linked to the adopted mass-loss rates in the underlying stellar and binary evolution models, especially for partially stripped envelopes \citep{Gilkis2022}. Finally, 87A-like SNe, which are those coming from H-rich blue supergiant progenitors, are found to be around $<1\%$ of all CCSNe, which is consistent with the observed rates of $<2\%$. 


\section{Supernovae interacting with circumstellar material produced through binary mass-transfer}
 Following the results from \cite{Ercolino2024,Ercolino2025}, if a binary system undergoes mass-transfer shortly before the SN, the unaccreted material may form nearby circumstellar material (CSM), which interacts with the SN ejecta shortly after the explosion. This may give rise to a so-called interacting supernova \citep{Smith2017}. Two classes of progenitor systems are identified: those in which the mass donor is a red supergiant in a wide binary undergoing Case C mass-transfer, which would produce an H-rich CSM and thus result in an interacting H-rich SN (\citealt{Ercolino2024}), and those in which the already binary-stripped donor star undergoes Case BB mass-transfer with the binary companion, producing a H-poor CSM and thus an interacting H-poor SN (\citealt{Ercolino2025}). In both channels, the progenitors are born with comparatively low initial masses ($M_\mathrm{i}\lesssim20\,M_\odot$).

 The population models find up to $2-4\%$ of all CCSNe to be H-rich interacting SNe, and $<4\%$ as H-poor interacting SNe. These numbers are consistent with the observed rate of Type~IIn and Type~Ibn SNe respectively (Fig.\,\ref{fig:SN_pop}). Moreover, the CSM that is produced is likely to be more focused along the orbital plane \citep{Lu2023} forming a disk-like structure, where this disk may remain bound \citep{Pejcha2016} and survive until the moment of the explosion of the SN progenitor \citep{Tuna2024, Chiba2026}. This agrees with observational evidence that many interacting SNe have aspherical CSM \citep{Soumagnac2020,Bilinski2024}. This asphericity is hard to reconcile with an eruptive mass-loss of a single star, and binary mass-loss offers a simple and natural explanation. These results suggest that binary evolution may produce the bulk of observed interacting SNe, although this scenario still fails to produce the outbursts observed in the years preceding many interacting SNe.

\section{Supernovae interacting with a binary companion}

\begin{figure}
    \centering
    \includegraphics[width=0.46\linewidth]{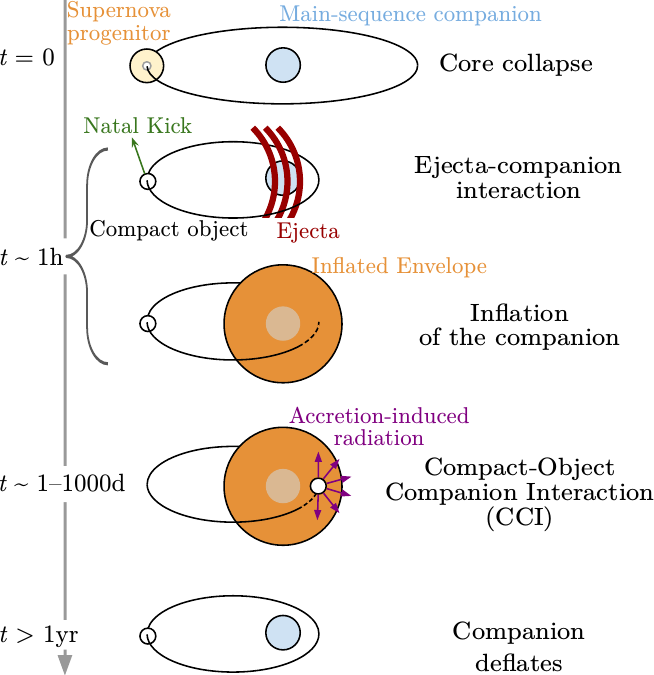}
    \hfill
    \includegraphics[width=0.50\linewidth]{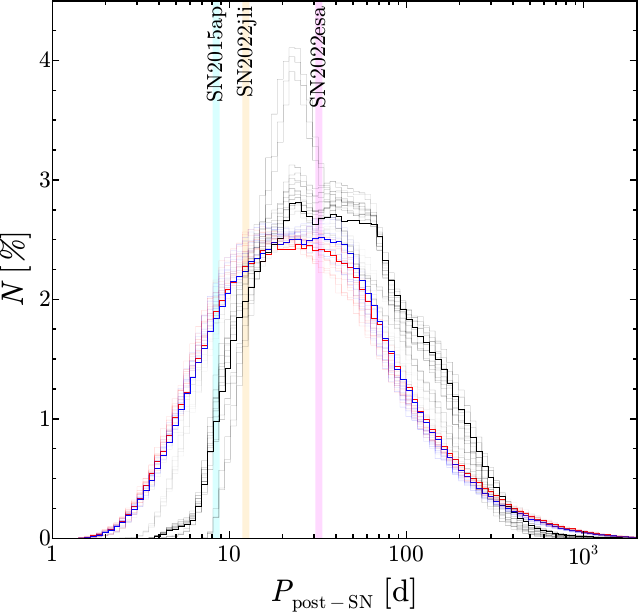}
    \caption{Left: Schematic evolution of close-by binaries following the explosion of the initially more massive companion, from core-collapse (top), through ejecta-companion interaction, and compact-object-companion interaction (CCI), to the deflation of the companion (bottom), with characteristic timescales from the time of core collapse. Right: post-SN orbital period distributions for binary models where periodic CCI is expected, where each individual distribution corresponds to a different population model, normalized to 100\%, and color-coded by the kick distribution adopted (red for \citealt{Disberg2025}, blue for \citealt{Kruckow18}, and black for \citealt{Valli2025}). The observed modulation periods of SN2022jli \citep{Moore2023,Chen2024}, SN2015ap \citep{Ragosta2025}, and SN2022esa \citep{Maeda2026} are also shown. Figures adapted from \cite{Ercolino2026b}.}
    \label{fig:CCI}
\end{figure}

Some binaries remain so close at the time of the first explosion that the companion star may be affected by interaction with the ejecta (left panel, Fig.\,\ref{fig:CCI}). When this occurs, the star may be thrown out of thermal equilibrium and inflate \citep[e.g.,][]{Hirai2018}. This expansion is taken into account in the population models, using the semi-analytical formulae from \cite{Hirai2018} and \cite{Ogata2021}.  If the natal kicks do not break up the binary, it is likely that the newly-born neutron star may skim the low density inflated envelope of the companion, emitting radiation as it sweeps it. If periodic, this interaction process, called companion--compact-object interaction (CCI), may produce periodic features in the SN \citep{Hirai2025}. 

The population models investigated show that periodic CCI is expected to be very rare in H-rich SNe, as their progenitors orbit a non-degenerate companion only in very wide binaries, which are easier to unbind with kicks. The opposite is true for H-poor SNe. Depending on the kick prescription, as well as the other model parameters explored, the fraction of H-poor SNe undergoing periodic CCI can range from 3\% to 27\%. The fraction of these models which would then exhibit an observable modulation in the light curve is yet to be determined.

The models provide an orbital period distribution for the binaries undergoing periodic CCI right after the explosion (right panel, Fig.\,\ref{fig:CCI}), which would manifest itself as the periodicity of the modulation of the signal from the resulting SN. The derived periods range from a few days to up to a few years, regardless of the population model used. This may be used as a bias to search for periodic behaviors in SN light curves.

So far, the only non-superluminous SNe observed with periodic undulations in their light curves are the H-poor transients SN2022jli \citep{Moore2023, Chen2024}, SN2015ap \citep{Ragosta2025}, and SN2022esa \citep{Maeda2026}. In particular, SN2022jli is the most convincing transient to have undergone periodic CCI \citep{Hirai2025}, as both light curve, spectra, and $\gamma$-ray flux all exhibit a $12.4\,\mathrm{d}$ modulation  \citep{Moore2023,Chen2024,Zhang2025}. Most of the population models investigated find individual binary evolution models that match the constraints of these transients. Because of that, the properties of the companion stars of these SNe can be estimated, like their luminosity and the duration of their inflated phase. SN2022jli is the most promising SN in which the companion is is still potentially observable, for example in the $J$-band by the Very Large Telescope's HAWK-I instrument \citep{GD2012, Ercolino2026b}.

\section{Summary and outlook}
Binary interaction is a dominant driver of the observed diversity of core-collapse supernovae. Population synthesis models based on comprehensive and large-scale grids of state-of-the-art single- and binary evolution models broadly reproduce the observed relative fractions of Type~IIP/L and Type~Ibc SNe, and imply that the majority of their progenitors have been affected by mass transfer, as donor stars, accretors, or even mergers. This is particularly evident in the ejecta mass distribution predicted for Type~Ibc supernovae, which offers a testable prediction with the results from future observations. The population models also predict that late-stage binary mass-transfer can potentially account for much of the observed population of interacting H-rich (Type~IIn) and H-poor (Type~Ibn) supernovae. This naturally explains the aspherical circumstellar medium inferred in many interacting supernovae, but it does not account for the pre-explosion outbursts often observed. Lastly, binaries that remain bound after the first supernova explosion could undergo periodic interaction between the compact-object and the companion star, which is estimated to affect up to $27\%$ of H-poor supernovae. This interaction process is potentially connected to periodic modulations in the SN light curve, as observed in SN2022jli, SN2015ap, and SN2022esa. Especially for SN2022jli, individual binary models matching its properties suggest that the companion could be observed. These results motivate the need for archival searches for past transients exhibiting periodic modulation, as well as to plan strategies to identify future events with the current and future observational facilities. 

Future supernova samples from current and future observational facilities (e.g. the Palomar Observatory, the Vera C. Rubin Observatory and the Argus Array) will continue the current revolution in transient discoveries, and are expected to yield improved estimates on the volume-limited rates and properties of different types of supernovae, as well as reveal populations of supernovae with periodic modulations. These results will be crucial testbeds to constrain the uncertain physics of single and binary evolution models.

\end{document}